# Leveraging Machine Learning to Gain Insights on Quantum Thermodynamic Entropy

Srinivasa Rao. P

∇×V Techno Labs, Hyderabad, India
`er.p.srinivas@gmail.com`

**Abstract.** The engine we are analyzing uses a single quantum particle as its working fluid, similar to Szilard's classical single-particle engine. Szilard resolved Maxwell's Second Law paradox by creating a physical demon to operate his engine. The design of our quantum engine is modelled after the classically-chaotic Szilard Map, which carries out a thermodynamic cycle involving measurement, thermal-energy extraction, and memory reset. Our analysis centers on studying the engine that utilizes a single quantum particle as its working fluid, which is similar in design to the classical single-particle engine constructed by Szilard. The proposed quantum engine follows the pattern of the classically-chaotic Szilard Map, which involves a thermodynamic cycle of measurement, thermal-energy extraction, and memory reset basing the focus is on investigating the thermodynamic costs associated with observing and controlling the particle, and comparing these costs in the quantum and classical limits. Through our study, we aim to shed light on the thermodynamic trade-offs that arise from Lindauer's Principle for information-processing-induced thermodynamic dissipation in both the quantum and classical regimes. We demonstrate that by using machine learning methods the energy analysis can be performed and the quantum engine can be simulated considering the Szilard engine based Second Law of Thermodynamics in its working condition. However, the quantum engine operates using significantly different mechanisms than its classical counterpart. In the classical implementation, the thermodynamics are determined by the process of measurement and erasure. In contrast, the cost of inserting partitions plays a critical role in the quantum implementation.

## 1    Introduction

### 1.1    Quantum Thermodynamic Nature

Research on quantum thermodynamic systems is a thriving field with significant implications in various areas, such as quantum computing, quantum communication, and energy conversion. [1] Furthermore, the field plays a critical role in comprehending the behavior of small systems, like those present in nanotechnology and quantum materials. James Clerk Maxwell, a renowned physicist, challenged the concept of irreversibility introduced by the second law of thermodynamics by conducting a thought experiment known as Maxwell's demon. [2][3] In the experiment, he posited that if there existed a demon or an entity with complete particle knowledge in a gas mixture, it could transfer fast-moving particles from cold to hot reservoirs, seemingly defying the very principle that governs them. [3] The significance of Maxwell's demon lies in its contribution to discovering how entropy and information are interconnected. Through the demon's capacity to manipulate the particles in a gas with the use of information, it has become clear that there is potential for relaxation of the second law's restrictions on energy exchange between a system and its surroundings.[4] The resulting conclusions have led to ground-breaking discoveries in terms of the relationship between entropy and information, marking an important milestone within the both fields *i.e* in the fundamental understanding of Thermodynamics and in the field of Computer science.[5] The original formulations of the second law of thermodynamics by Clausius, Kelvin, or Planck did not include any reference to information, however in a more subtle way as a knowledge of the demon.[6][9] There are two challenges particularly in this case, one is to develop a detailed version of the second law that explicitly includes the role of information, besides it is necessary to establish a clear understanding of the physical properties of information, so that it can be incorporated into the second law in a concrete way rather than being treated as an abstract concept. [7][11] It is important to establish a clear understanding of the physical aspect of information, so that it can be explicitly incorporated into the second law as a tangible entity rather than an abstract concept. This would allow for information manipulations, such as measurement, erasure, copying, and feedback, to be considered as physical processes that entail thermodynamic expenses.[8][9][14]

A quantum thermodynamic system is a physical system that combines the principles of quantum mechanics and thermodynamics. It is a system that can undergo changes in energy, work, and entropy, while being described by quantum mechanical states and operations. A quantum thermodynamic system is characterized by particles whose energy levels are quantized and discrete, and whose interactions are governed by quantum mechanics. [10][11] The system's thermodynamic attributes, including its heat capacity, entropy, and temperature, are determined by the quantum states and actions of the particles within the system. Although a complete and precise understanding of how an engine works necessitates consideration of its fundamental dynamics, we focus on two key aspects of a quantum Szilard engine. [13][15] One is to recognize the significance of the information processing and provide a thorough, detailed analysis of the "demon" by endowing it with a physical representation. Furthermore, improve the explanation of the expenses incurred in each stage of the engine's thermodynamic cycle, which includes the measurement, thermal-energy extraction, and reset phases using computational setup.

### 1.2 The Design of Quantum Engine

The Quantum Szilard Engine (QSE) is an appropriate system for investigating the importance of information processing during thermodynamic changes. The engine is made up of two parts: the system under investigation (SUS) and a quantum demon or controller. These components are surrounded by an incoherent environment that keeps the composite system thermally balanced at an inverse temperature.

To begin, the system is defined, which includes the number of particles and the parameters that govern their behavior, such as energy levels and interactions, a protocol is created to control the dynamics of the system and extract useful work from it. This protocol typically entails splitting the system into two parts and controlling particle movement between these parts with external operations. The cost of includes the implementation of the protocol in place, in terms of energy and other resources, which is frequently accomplished by employing Lindauer's principle, which relates the amount of work required to erase a bit of information to the system's entropy increase. And finally, the performance of the engine is analyzed, including its efficiency, power output, and other relevant metrics. Optimization techniques, such as numerical simulations or machine learning algorithms, may be used to find the optimal parameters for the engine's design. To provide context for the analysis, it is necessary to first grasp the fundamentals of a quantum particle's thermodynamics in a one-dimensional box of length L. This system is frequently used as a simple model to study the behavior of confined particles in fields such as materials science and quantum mechanics.

## 2 Mathematical Modeling

The particle in a box system is defined by the size of the box and the particle's energy, which determine the entropy, temperature, and heat capacity of the system. Because of the confinement, the particle's energy is quantized, and the number of energy levels increases with the size of the box. As a result, the thermodynamic properties of the system are highly dependent on the size of the box and the energy of the particle. Various analytical and numerical methods have been developed to calculate the thermodynamic properties of a particle in a box system. Solving the Schrödinger equation for the particle's wave function and applying statistical mechanics to determine the thermodynamic properties of the system are two of these methods. However, due to the system's complexity, accurate calculations can be difficult for large or complex systems. As a result, machine learning models for predicting the thermodynamic properties of particles in a box system have been developed. These models use data from simulations and experiments to learn the relationship between the input parameters and the thermodynamic properties of the system, providing a new tool for understanding system behavior.

$$E_n = \frac{n^2 \pi^2 \hbar^2}{2ml^2} \qquad (1)$$

where n is a positive integer, m is the mass of the particle, and $\hbar$ is the reduced Planck constant.

The partition function, which gives the probability of finding the system in a given state at temperature T, is given by

$$Z = \sum_{n=1}^{\infty} e^{-\beta E_n} \tag{2}$$

$\beta = {1}/{k_B T}$ where $k_B$ is Boltzmann constant and T is the temperature

$$\langle E \rangle = \frac{\partial \ln(Z)}{\partial \beta} = \frac{\pi^2 \hbar^2}{6ml^2} \coth\left(\frac{\pi^2 \hbar^2}{2ml^2 k_B T}\right) \tag{3}$$

The heat capacity of the particle is given by

$$C_V = \frac{\partial \langle E \rangle}{\partial T} = \frac{\pi^2 \hbar^2}{3 k_B m l^2} \operatorname{csch}^2\left(\frac{\pi^2 \hbar^2}{2ml^2 k_B T}\right) \tag{4}$$

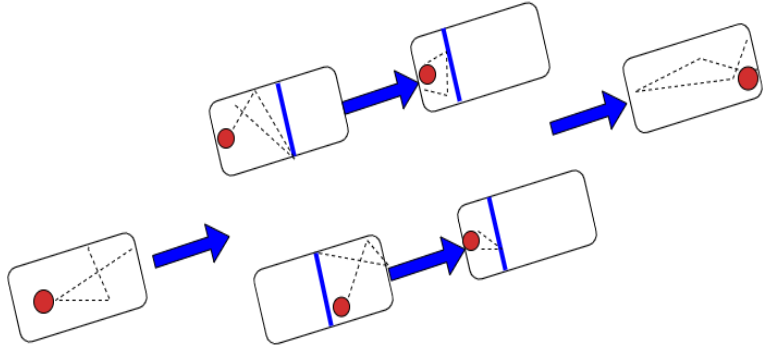

**Fig. 1.** Illustration of electron in a box and the movement of the piston.

The function is defined in such a way that it is set up the initial conditions for the simulation and partitions the box into two parts ($x_A$ and $x_B$) based on the particle positions, then perform the quasistatic compression/expansion by adjusting the particle momenta according to the position within the box. After the compression/expansion, the function calculates the work done by the engine using the final momenta of the particles and plots the final particle positions and momenta. Finally, the function returns the work done by the engine in Joules.

The simulation of a quasistatic Szilard engine is defined as a theoretical engine that uses information to extract work from a system, then it is modeled as a harmonic oscillator in contact with a heat bath at a constant temperature. the work done by the engine would be calculated and plotted as the probability distribution of the position of the piston in the engine. However, the histogram of position as shown in the figure 2, which is obtained from the probability distribution which shows the frequency of occurrence of different piston positions, which gives an indication of the stability of the system at those positions.

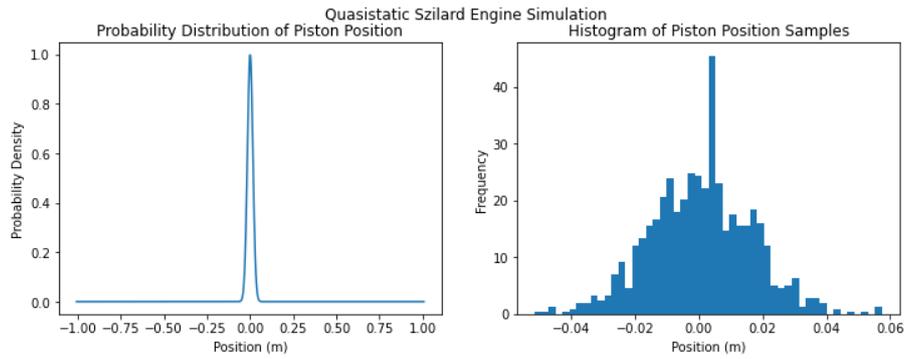

**Fig. 2.** Quasistatic Szilard Engine normalized for time.

The distribution can be understood easily after normalization as in figure 3, which gives a very clear indication about the piston stability of the system and its operational accuracy.

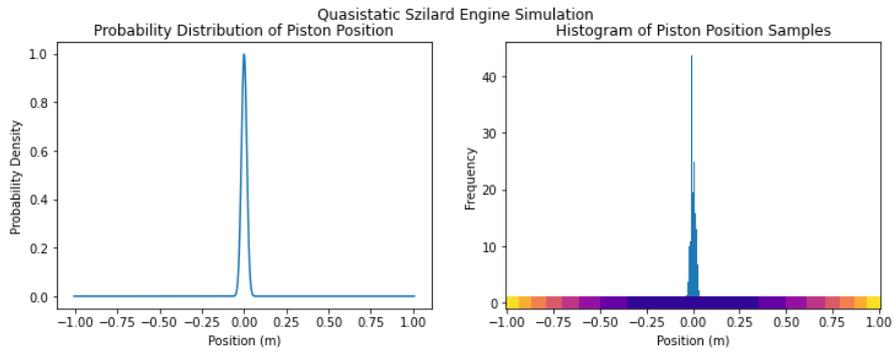

**Fig. 3.** Quasistatic Szilard Engine normalized for time.

As shown in the histogram mentioned as shown in the figure.4 the narrow and centered around a particular position suggests that the piston is stable at that position and that the engine is operating efficiently.

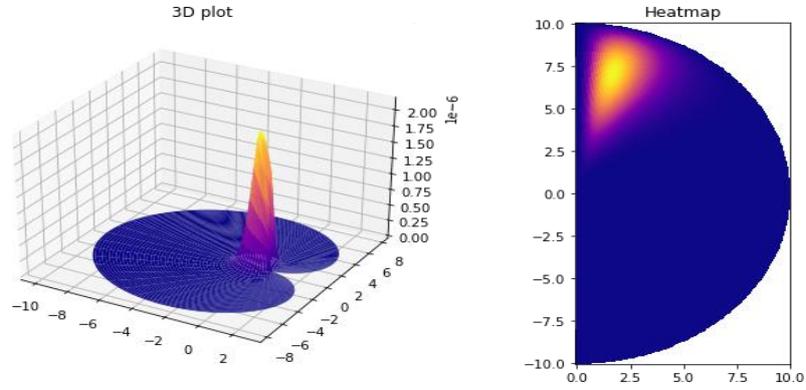

**Fig. 4.** Histogram representation of the frequency.

It has been used as a model system for studying the thermodynamics of information processing and the limits of information processing imposed by the laws of thermodynamics. The engine has also been used to study the role of information in the Maxwell's demon thought experiment, which involves a demon that uses information to violate the second law of thermodynamics by selectively opening and closing a partition between two gases of different temperatures. By examining the residual plot for thermodynamic entropy in the Szilard engine model and after simulation of the model using machine learning algorithm, it is possible to identify any systematic errors or biases in the model.

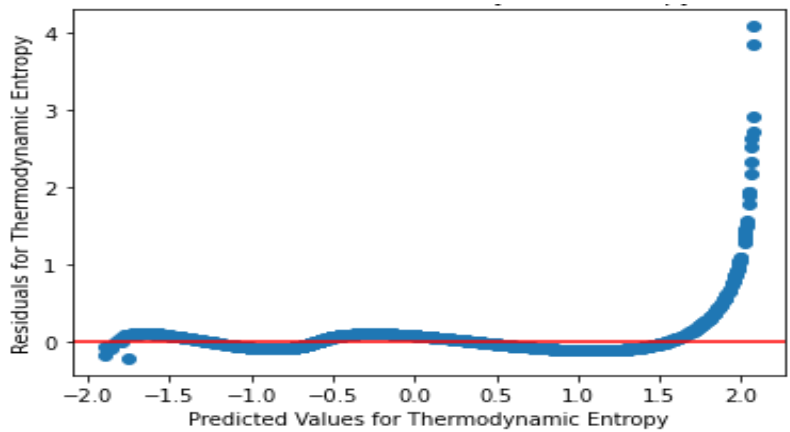

**Fig. 5.** Thermodynamic Entropy.

This can help to refine the model and improve the accuracy of the simulation results. Additionally, residual plots can be used to compare the performance of different simulation models, helping to identify the most accurate and reliable approach to modeling the Szilard engine and trace the nature of entropy variation contrasted along the residual thermodynamic entropy and predicted values as shown in the fig. 5. A good residual plot would show how the random scatter of data points around zero, indicating that the model is a good fit for the data. It is found that there is a pattern in the residual plot which indicates that the model may not be capturing all the underlying factors that contribute to the change in entropy over time. It is important to analyze the residual plot along with other statistical metrics R-squared and the p-value to fully understand the performance of the simulation model in predicting the change in entropy over time as shown in fig.6.

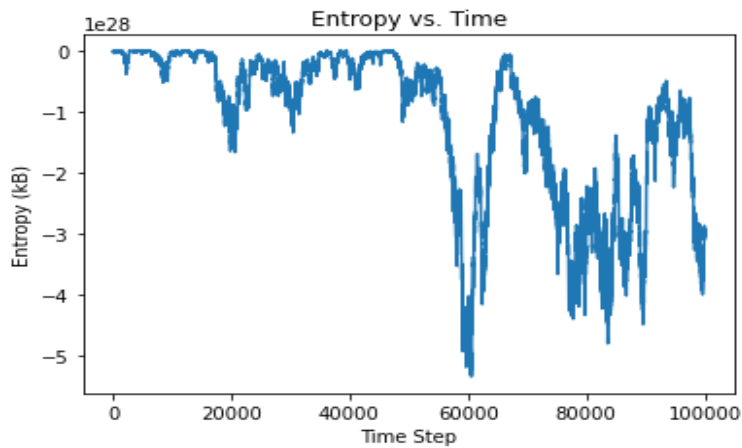

**Fig. 6.** Thermodynamic Entropy Vs Time

The engine optimization using machine learning to analyze and optimize the performance of the engine. which be used to predict the behavior of the engine under different conditions and to identify optimal operating parameters, such as the temperature and pressure of the working fluid.

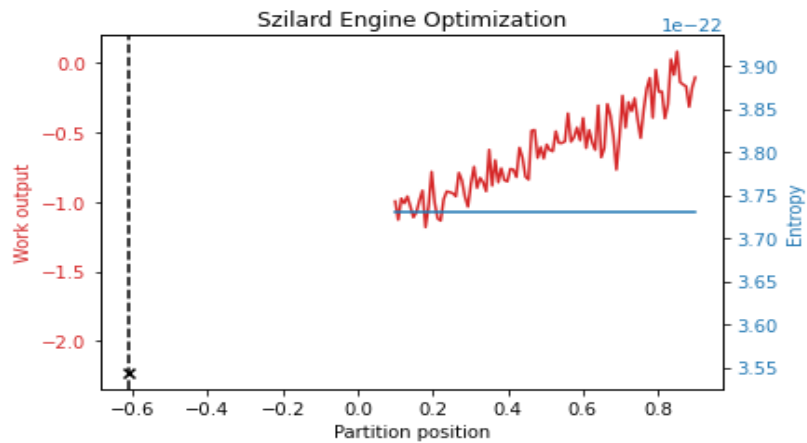

**Fig. 7.** Szilard Engine optimization using reinforcement learning

This has led to improved efficiency and performance of the engine, as well as reduced emissions and energy consumption. The simulation using neural networks is performed and trained on data to predict the performance of the engine under different conditions, while reinforcement learning technique is used to optimize the engine's control strategy which is shown in the figure.7.

## Conclusion

The work presents an analysis of a quantum engine that uses a single quantum particle as its working fluid, based on Szilard's classical single-particle engine. The design of the proposed quantum engine follows the pattern of the classically-chaotic Szilard Map, involving a thermodynamic cycle of measurement, thermal-energy extraction, and memory reset. It is also focused on the thermodynamic energy associated with the particle in a box, and comparing the same of the quantum and classical regimes, using machine learning methods. It is important to note that the quantum engine operates using significantly different mechanisms than its classical counterpart, where the cost of inserting partitions plays a critical role in the quantum implementation incorporating maxwell's demon. The study sheds light on the thermodynamic trade-offs that arise from Lindauer's Principle for information-processing-induced thermodynamic dissipation in both the quantum and classical regimes and how the machine learning approach could help in understanding the concept from the fundamental perspective.